\documentclass[10pt,twocolumn, nofootinbib]{revtex4-2}
\usepackage{amsmath}
\usepackage{amssymb}
\usepackage{graphicx}
\usepackage{amsfonts}
\usepackage{enumitem}
\usepackage{graphicx}
\usepackage{hyperref}
\usepackage{tikz}

\hypersetup{
	colorlinks=true,
	citecolor=blue,
	urlcolor=blue,
	linkcolor=blue
}
\begin{document}

\title{On the reality of the quantum state once again: A no-go theorem for $\psi$-ontic models}
\author{Gabriele Carcassi}
\affiliation{Physics Department, University of Michigan, Ann Arbor, MI 48109}
\author{Andrea Oldofredi}
\affiliation{Centre of Philosophy, University of Lisbon, Portugal}
\author{Christine A. Aidala}
\affiliation{Physics Department, University of Michigan, Ann Arbor, MI 48109}
\vspace{2mm}

\date{\today}

\begin{abstract}
In this paper we show that $\psi$-ontic models, as defined by Harrigan and Spekkens (HS), cannot reproduce quantum theory. Instead of focusing on probability, we use information theoretic considerations to show that all pure states of $\psi$-ontic models must be orthogonal to each other, in clear violation of quantum mechanics. Given that (i) Pusey, Barrett and Rudolph (PBR) previously showed that $\psi$-epistemic models, as defined by HS, also contradict quantum mechanics, and (ii) the HS categorization is exhausted by these two types of models, we conclude that the HS categorization itself is problematic as it leaves no space for models that can reproduce quantum theory.
\end{abstract}

\maketitle

\section{Introduction}
 
In 2010 Harrigan and Spekkens (HS) proposed a formal classification in order to categorize the nature of the quantum state, i.e.\ to establish whether in a certain model $\psi$ corresponds to a real property of a quantum object, in which case the model is called $\psi$-ontic, or to some observer information, making it $\psi$-epistemic \cite{Harrigan:2010}. While their original aim was to clarify Einstein's view of Quantum Mechanics (QM), the HS framework has been widely employed in the literature not only to categorize different formulations of QM, but also to argue what types of interpretations are admissible (\cite{Leifer:2013, Leifer:2017, Branciard:2014, Hermens:2021, Wood:2015, Ringbauer:2015, Mazurek:2016, Bartlett:2012}; cf.\ \cite{Oldofredi:2020b, Ladyman:2021} for critical discussions). 

Referring to this, one of the most influential works based on HS classification is due to Pusey, Barrett and Rudolph who published a formal result in \emph{Nature Physics}---widely known as the PBR theorem---showing that ``if the quantum state merely represents information about the real physical state of a system, then experimental predictions are obtained that contradict those of quantum theory'' (\cite{PBR:2012}, p.\ 475). Alternatively stated, PBR argued that in every model reproducing the statistics and predictions of QM the quantum state $\psi$ must represent real physical properties of the system under consideration and not agents' knowledge---i.e.\ models must be $\psi$-ontic. Consequently, quantum theories cannot be $\psi$-epistemic. 

Such a theorem had a remarkable resonance \cite{Leifer:2014, Leifer:2014b, Lewis:2012, Renner:2012, Colbeck:2017, Hardy:2013, Maroney:2014, Patra:2013, Mansfield:2016, Schlosshauer:2012, Schlosshauer:2013, Schlosshauer:2014, Aaronson:2013}, and questions about its actual meaning are still discussed today: on the one hand, some authors believe that it rules out interpretations of QM where $\psi$ merely represents information. On the other hand, it has recently been shown by other scholars that non-trivial epistemic as well as statistical approaches to QM are not refuted by the PBR argument \cite{Ben:2017, Rizzi:2018, Oldofredi:2020b, Oldofredi:2021, DeBrota:2019}.

While the discussion has been focused mainly on whether $\psi$-epistemic models are problematic or whether the PBR argument itself is problematic, we ask a different question: what if the underlying HS classification itself is problematic? In particular, the framework assumes that once pure states are fully characterized, mixed states are constructed simply as a standard (i.e.\ classical) statistical mixture. However, as it is well-known, classical and quantum mixtures have different properties. For instance, classical mixtures have a single decomposition in terms of pure states while quantum mixtures do not. Using information theoretic considerations on mixtures, then, provides a new approach to explore possible violations of quantum mechanics.

Employing this new approach, in this paper we show that $\psi$-ontic models cannot reproduce all the predictions, results and implications of QM. We start by noting that the von Neumann entropy plays a crucial role in the predictions of both quantum statistical mechanics and quantum information theory. We show that the degree of overlap between two probability distributions affects how the entropy of their mixture relates to the entropy of the components. We find that the lack of overlap of epistemic states---a necessary condition for a model to be $\psi$-ontic---requires that all pure states must be orthogonal, which directly contradicts quantum theory.

Combining our result with the PBR argument, we conclude that the HS classification itself is fundamentally flawed: both $\psi$-epistemic and $\psi$-ontic models contradict quantum mechanics.\footnote{In principle, PBR leaves open the possibility of $\psi$-epistemic models that violate statistical independence. Given that statistical independence is linked to entropy additivity, we will see that that loophole can be closed as well.} Consequently, the HS framework should be employed neither to classify interpretations of the quantum formalism, nor to draw conclusions about the nature of $\psi$. 

\section{Summary of the Harrigan \& Spekkens Model}

We briefly review the main features of the classification provided by HS starting with the usual operational setting for QM. We have a preparation protocol $P$, associated with a density operator $\rho$ on the relevant Hilbert space, and a measurement protocol $M$, represented by a POVM $\{ E_k\}$ where each $k$ represents a possible measurement outcome. The probability of obtaining a particular $k$ given a particular preparation $P$ and measurement $M$ is given by the generalized Born rule
\begin{equation}
	p(k|M, P)=\textrm{tr}(\rho E_k).
\end{equation}

An ontological model, as defined by HS, additionally assumes that there exists a set of states $\Lambda$, called \textbf{ontic states}, that provide the complete specification of the properties of a given physical object. A preparation $P$ will prepare a particular ontic state $\lambda$ according to the probability distribution $p(\lambda | P)$, which is referred to as \textbf{epistemic state}. The measurement outcome will depend only on the ontic state with probability $p(k|\lambda, M)$ (\cite{Harrigan:2010}, p.\ 128). This leads to the following expression:
\begin{equation}\label{prob_condition}
	\int_\Lambda d\lambda p(k|\lambda, M) p(\lambda| P)= \textrm{tr}(\rho E_k).
\end{equation}
It is also assumed that a mixture of pure states $\{ \psi_i \}$ with probabilities $\{ w_i \}$ will be represented by
\begin{equation}\label{epistemic_mixing}
	\sum_i  w_i p(\lambda| P_{\psi_i}).
\end{equation}
As the epistemic states for mixed preparations are simply linear combinations of epistemic states for pure preparations, the HS model concentrates on the latter.

\begin{figure}
\includegraphics[scale=.7]{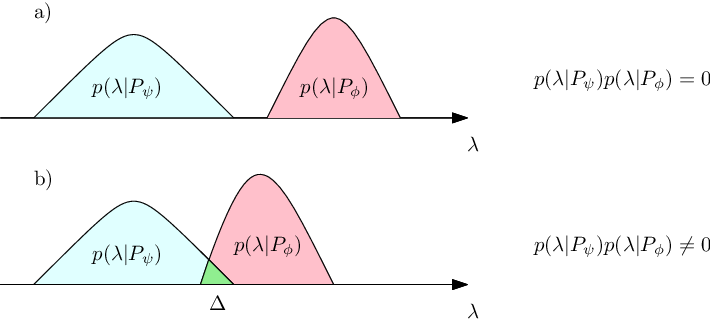}
\caption{\footnotesize{Harrigan and Spekkens' distinction between $\psi$-ontic (a) and $\psi$-epistemic (b) ontological models.}}\label{overlap}
\end{figure}

To classify an ontological model, we look at the relationship between quantum and ontic states. There are two broad categories. In a \textbf{$\psi$-ontic} model the wave function is a physical property of the ontic state, in the sense that given an ontic state $\lambda$ there is only one pure state preparation $P_\psi$ that could have prepared $\lambda$. As we can see in figure \ref{overlap} (a), this happens if the probability distributions do not overlap, i.e.\ if we have
\begin{equation}\label{ontic_condition}
	p(\lambda | P_{\psi})p(\lambda|P_{\phi})=0
\end{equation}
for all pairs of states $\psi$ and $\phi$. If a model is not $\psi$-ontic, then it is \textbf{$\psi$-epistemic}. In this case, $\lambda$ can be described by more than one quantum state and the wave function is taken to represent knowledge about the state preparation---in such models quantum states generate overlapping probability distributions over $\Lambda$ as shown in figure \ref{overlap} (b). A $\psi$-ontic model is said \textbf{$\psi$-complete} if the quantum states and the ontic states coincide. More precisely if 
\begin{equation}\label{complete_condition}
	p(\lambda|P_\psi)=\delta(\lambda-\lambda_{\psi}).
\end{equation}
All other models are \textbf{$\psi$-incomplete}. For specific examples of ontological models, see \cite{Harrigan:2010}, sections 2.4.1--3.

To make this classification more concrete, let us give an example from classical mechanics. Consider the case where we prepare a particle according to a specific value of energy. The energy partitions phase space into mutually exclusive regions, and therefore we can understand the energy of the preparation as a property of the particle itself. According to HS, this would be an \emph{ontic property}. Consider now the case where we prepare a particle according to a specific temperature. When we take the particle from the oven, we are sampling from a Boltzmann distribution over different energies. However, unlike energy, temperature does not partition phase space because the same particle state could have been prepared by ovens at different temperature. Temperature is a property of the preparation and therefore an \emph{epistemic property}. 

After having introduced the relevant definitions of the HS ontological model framework, it should be noted that it has been subject to several criticisms, as already said in the previous section. For instance, Oldofredi \& Lopez in \cite{Oldofredi:2020b} analyzed the implicit assumptions made by Harrigan \& Spekkens in order to define their categorization, and showed that relevant interpretations of quantum theory---such as the statistical, relational and perspectival readings of the quantum formalism---cannot be classified according to the HS approach, given that their principles are in tension with those employed in \cite{Harrigan:2010}. Similarly, Hance, Rarity and Ladyman argued in \cite{Ladyman:2021} that the dichotomy ``ontic vs.\ epistemic'' proposed by HS is artificial, since there are cases in which the quantum state can represent both ontic properties of quantum systems as well as agents' knowledge of them. Thus, they claim, the definitions of $\psi$-ontic and $\psi$-epistemic introduced a few lines above do not capture the intuitive ideas with which physicists and philosophers alike use the terms ``ontic'' and ``epistemic''.

While we agree with the main messages of these critical responses to the HS framework, the argument offered in this paper follows a different and complementary strategy. In fact, instead of criticizing the assumptions used by Harrigan \& Spekkens, we consider them valid (for the sake of the argument) and derive a no-go theorem for $\psi$-ontic models directly from them, as we are going to show in the next section.

\section{Entropy and the ontological model}

Looking at the HS classification it may seem that one has complete freedom in choosing the overlap between epistemic states. In this section we will show that this is not the case: this choice is in fact constrained by entropy of mixtures. Moreover, these constraints cannot be replicated by a $\psi$-ontic model with standard (i.e.\ classical) information theoretic techniques. Since quantum statistical mechanics and quantum information theory depend on the correct value of entropy on mixtures, $\psi$-ontic models will violate quantum theory.

Let us denote $M(\Lambda)$ the space of all probability distributions (i.e. all probability measures) over the space of ontic states $\Lambda$. Let us denote $\mathcal{H}$ the Hilbert space of the corresponding quantum system and $M(\mathcal{H})$ the space of all mixed states (i.e. the space of positive semi-definite trace-one Hermitian operators). An ontological model, then, must give us an appropriate $\rho(\lambda) \in M(\Lambda)$ for every $\rho \in M(\mathcal{H})$ such that all predictions are satisfied. This includes those given by statistical mechanics and information theory, which depend on the von Neumann entropy $H_\mathcal{H} : M(\mathcal{H}) \to \mathbb{R}$. Given that $M(\Lambda)$ is the space of probability measures over $\Lambda$, it is natural to use the Shannon/Gibbs entropy function $H_\Lambda : M(\Lambda) \to \mathbb{R}$, as this is what is typically done in statistical mechanics.\footnote{Note that, while many functions called ``entropy'' exist, the Shannon entropy is the only indicator of variability that is continuous, monotonic and linear in probability \cite{Carcassi:2021}. It is also the one that provides the correct link to the thermodynamic entropy.} Since we want to be able to replicate the predictions of statistical mechanics and information theory, a very reasonable request is that these two entropies, calculated on the different representations, agree.

We now introduce two lemmas that show how the entropy behaves when mixing two states, both for the Shannon/Gibbs entropy $H_\Lambda$ and the von Neumann entropy $H_\mathcal{H}$.\footnote{Full calculations with all steps are included in the appendix.} The first will be used to calculate entropy of statistical mixtures of epistemic states, since equation \ref{epistemic_mixing} tells us the mixture is done according to a standard (classical) measure of probability. The second will be used to calculate the entropy on the quantum case.

\textbf{Information entropy of mixed non-overlapping probability distributions}. Let $\rho_1, \rho_2 \in M(\Lambda)$ be two probability density functions over a space $\Lambda$ with measure $\lambda$.\footnote{The notation of HS uses $\lambda$ as both the value and measure for integration, as often done in single variable integrals.} That is, they are the Radon-Nikodym derivatives of the respective probability measures with respect to the measure $\lambda$. The entropy for each distribution is given by the usual formula\footnote{The logarithm is assumed to be in base 2.}, for example
\begin{equation}\label{shannon_entropy}
	H_\Lambda(\rho_1) = - \int_\Lambda \rho_1(\lambda) \log \rho_1(\lambda) d\lambda.
\end{equation}
Suppose the two distributions are disjoint, and let $\rho = \frac{1}{2} \rho_1 + \frac{1}{2} \rho_2$ be a uniform mixture of the two distributions. More precisely, $\rho$ is the Radon-Nikodym derivative of the average of the two probability measures with respect to $\lambda$. The entropy of $\rho$ is given by
\begin{equation}\label{entropy_nonoverlap}
	H_\Lambda(\rho) = 1 + \frac{1}{2} H_\Lambda(\rho_1) + \frac{1}{2} H_\Lambda(\rho_2).
\end{equation}
Note how the non-overlapping assumption fixes the entropy of the mixed state. 

\textbf{Quantum information entropy of quantum mixed states}. Now suppose $\psi$ and $\phi$ are two pure quantum states and let $p = | \langle \psi | \phi \rangle |^2$ be the probability of transition from one to the other. Consider the mixed state $\rho = \frac{1}{2} | \psi \rangle \langle \psi | + \frac{1}{2} | \phi \rangle \langle \phi |$. Its entropy is given by
\begin{equation}\label{entropy_mixed}
	\begin{aligned}
	H_\mathcal{H}(\rho) &= H_S\left(\frac{1+\sqrt{p}}{2}, \frac{1-\sqrt{p}}{2}\right) \\
	&= - \frac{1+\sqrt{p}}{2} \log \frac{1+\sqrt{p}}{2} - \frac{1-\sqrt{p}}{2} \log \frac{1-\sqrt{p}}{2} ,
	\end{aligned}
\end{equation}
where the right side of the equation is the Shannon entropy calculated on the given values.\footnote{Note that $H_S$ is the ``discrete'' Shannon entropy that acts on a countable set of coefficients that sum to one.} Therefore the entropy of an equal mixture of two pure states depends only on the probability of measuring one having prepared the other.

\textbf{Theorem: Since non-overlapping distributions can only represent orthogonal states, $\psi$-ontic models cannot be consistent with quantum theory.}\footnote{Note that the theorem applies to all $\psi$-ontic models, including the $\psi$-supplemented ones.}

\emph{Proof}: Suppose we have a $\psi$-ontic model defined according to \cite{Harrigan:2010}. The epistemic states $p(\lambda|P_\psi)$ and $p(\lambda|P_\phi)$ consist of non-overlapping probability distributions over a space $\Lambda$. That is, they are two non-overlapping probability density functions defined to be the Radon-Nikodym derivatives between the respective probability measures over $\Lambda$ and the measure $\lambda$. Equation \ref{epistemic_mixing} tells us that the probability measures combine in the usual way, and therefore an equal mixture of the two states must obey equation \ref{entropy_nonoverlap}. Given that $\psi$ and $\phi$ are pure states, and the entropy for pure states in quantum mechanics is zero, we must have
\begin{equation}\label{entropy_pure}
	H_\Lambda(p(\lambda|P_\psi)) = H_\Lambda(p(\lambda|P_\phi)) = 0,
\end{equation}
and therefore
\begin{equation}\label{required_entropy}
	\begin{aligned}
		H_\Lambda\left(p(\lambda|\frac{1}{2}P_\psi + \frac{1}{2}P_\phi)\right) &= \\
		H_\Lambda\left(\frac{1}{2}p(\lambda|P_\psi) + \frac{1}{2}p(\lambda|P_\phi)\right) 
		&= 1.
	\end{aligned}
\end{equation}
If we compare the above with eq.~\ref{entropy_mixed}, it follows that $p$ must be zero. In fact, the only case in which the Shannon entropy for a two-element distribution is equal to one is for the uniform distribution $H_S(1/2, 1/2) = 1$.  That is, we must have that
\begin{equation}\label{orthogonal}
	\langle \psi | \phi \rangle = 0
\end{equation}
no matter what $\psi$ and $\phi$ are.

Hence, the non-overlapping assumption built into the $\psi$-ontic model necessarily implies that all pure states are orthogonal. Since this is not true in quantum mechanics, any $\psi$-ontic model will fail to reproduce the results of quantum information, quantum statistical mechanics and, therefore, quantum theory in general. $\square$

\section{Throwing out information theory}

Note that our conclusion relies on the use of standard information theoretic tools. In principle, this does not exclude that someone may find a different construction to reproduce the correct entropies. That is, one may give a different definition of entropy to use on the ontic space. We believe this to be highly unlikely as there are, at least, two objective difficulties that cannot be circumvented.

First of all, information entropy, in all its versions, is strictly concave, meaning that $H(p \rho_1 + (1-p) \rho_2) \geq p H(\rho_1) + (1-p) H(\rho_2)$ with the equality valid only if $\rho_1$ and $\rho_2$ are the same exact state. That is, we can check whether two states $\rho_1$ and $\rho_2$ are the same state simply by creating a mixture $\rho = \frac{1}{2} \rho_1 + \frac{1}{2} \rho_2$ and see whether we end up with the same entropy or not. In other words: \emph{entropy fixes equality}.
The issue here is that the mapping between epistemic states and density matrices is not injective. Quantum mixtures, in fact, do not have a single decomposition in pure states. For example, the maximally mixed state of a qubit for a spin-$1/2$ system can be achieved with either an equal mixture of spin up and down or an equal mixture of spin left and right. Therefore epistemic states cannot be mapped one-to-one to quantum density operators, but rather each density operator corresponds to an equivalence class of epistemic states.\footnote{The degree of degeneracy one has to require is also extreme. For a qubit, in the simplest case, we would have a $\psi$-ontic state for each direction on the Bloch sphere, meaning the space of ontic states is the sphere $S^2$. The space of all possible epistemic states is the set of all possible distributions integrable to one. This is an infinite dimensional function space. However, the space of mixed states for a quibit is the interior of the Bloch sphere, which is a bounded three-dimensional manifold. The equivalence classes of epistemic states, then, must be infinite dimensional.} This degeneracy means that the notion of equality among epistemic states is different than the notion of equality among density operators. In turn, this means that whatever definition we have for entropy on epistemic states must not satisfy strict concavity, if we want the value of entropy to agree between epistemic states and density operators. \emph{Entropy cannot fix equality for epistemic states}. In other words, one can make mixtures of different epistemic states without increasing entropy, which is questionable on physics grounds.

Secondly, as equation \ref{entropy_mixed} shows, computing the entropy requires the inner product. In fact, if we had the entropy for all uniform mixtures of all pairs of pure states, we could reconstruct the inner product. Therefore, whatever other structure one puts on top of epistemic states essentially redefines the inner product. Also, it has to do it in a way compatible with the other definitions of the model, therefore this structure will likely be ad-hoc for each model. In other words, the idea that the inner product just represents transition probabilities, which, in our view, is the key idea the ontological model uses to explain ``what really happens'' does not work.

In other words, the use of standard probability measures to describe the mixing is problematic in the same way that the use of standard probability measures would be a problem to describe the probability of transition. If we step back, we can understand our result and the PBR result in a broader context.

\section{The problem with ontological models}

If we combine our result with the PBR theorem, the only HS ontological models that are not ruled out are $\psi$-epistemic models in which quantum objects are not independently prepared. That is, where the joint distribution over ontic states is not the product of the marginals. However, this loophole can also be closed with information theoretic considerations. It is a well-known result in both classical and quantum information theory that the entropy of a joint distribution is the sum of the marginal if and only if the subsystems under consideration are independent (\cite{Ash:2010}, \cite{Nielsen:2010}). Therefore an epistemic state that represents independent quantum states must also be the joint distribution of statistically independent epistemic states of the individual systems. This tells us that $\psi$-epistemic models must represent composite states of independently prepared systems with independently distributed epistemic states, and therefore PBR applies. This closes the last loophole: no model in the HS categorization can reproduce quantum theory in its entirety, which includes quantum information and quantum statistical mechanics.

Thus, one concludes the following:
\begin{quote}
	\textbf{No ontological model can reproduce quantum theory}.
\end{quote}
Although this result may at first sight seem paradoxical, once we analyze in detail the mathematics behind HS categorization we see that both results---ours and PBR---are finding something, in retrospect, self-evident.

It is well established that classical probability theory cannot recover the probability of quantum transitions. So, how does the HS model try to accommodate this? Given that $p(\lambda|P)$ combines using classical probability theory, all the ``quantumness'' must be in $p(k|\lambda, M)$. In fact, note that this second expression does not allow us to write the joint probability distribution for multiple observers; therefore, each ontic state is not partitioned further. Therefore, if $\lambda$ is the state of a single quantum system, we can ``hide'' the inner product inside $p(k|\lambda, M)$; this is enough to reproduce quantum probability for pure states of single quantum systems, but not much more.

In our result, we saw that the use of standard measure theory when creating mixtures as in equation \ref{epistemic_mixing} generates a problem because the convex space of classical distributions is significantly different from the convex space of quantum mixed states. For this case, ontological models are ``not quantum enough''.

In PBR, the issue is breaking a composite system into parts. They find  that if the probability measure on $\lambda$ is factorized, that is (\cite{PBR:2012}, eq. 5) 
\begin{equation}
	\int_{\Lambda} \dots \int_{\Lambda} p(k | \lambda_1, \cdots, \lambda_n) \mu_{x_1}(\lambda_1)\cdots\mu_{x_n}(\lambda_n) d\lambda_1 \cdots d\lambda_n ,
\end{equation}
there is a problem. However, the issue here is not the factorization, but rather breaking the composite system into parts, each with its own ontic state as in 
\begin{equation}
	\begin{aligned}
	\int_\Lambda &p(k | \lambda) d\lambda = \\
	&\int_{\Lambda_1} \dots \int_{\Lambda_n} p(k | \lambda_1, \cdots, \lambda_n) \mu(\lambda_1, \cdots, \lambda_n) d\lambda_1 \cdots d\lambda_n.
	\end{aligned}
\end{equation}
This assumes that a composite system is describable using a standard classical probability measure, which, in turn, allows us to express the joint probability distribution of observables of different subsystems. Again, this in general will fail because it is ``not quantum enough''.

Therefore there are two different ways ontological models use standard measure theory: one in mixtures, highlighted by us; one in composite systems, highlighted by PBR. The use of measure theory is what drives the failure of the model.

Now, the fact that quantum theory does not follow the rules of classical-Kolmogorov probability is nothing new \cite{Werner:2014, Pitowsky:1989}. However, this does not seem to be enough to clear confusion on the subject because, admittedly, we do use classical probability in the context of quantum mechanics. For example, we can imagine preparing the direction of spin based on a classical distribution. Furthermore, the output of a measurement is described by classical probability. The temptation, then, is to simply assume that all we need to do is just put the right transition probability between the two spaces, and we are done. This is exactly what cannot be done and what does not work. That is
\begin{quote}
	\textbf{In the context of quantum mechanics, standard probability measures are allowed over preparations and measurement outcomes, but not over states.}
\end{quote}
In our approach, we exploited the fact that the entropy calculation uses the geometry of the inner product without invoking a probability of transition. That is, two pure states are orthogonal not because one cannot measure the first having prepared the second, but because their equal mixture raises the entropy by one bit. Therefore quantum mechanics is not simply defining a probability of transition between preparations and measurements: it is doing something more. While it is not in the scope of this article to articulate precisely what this ``more'' is, which will be the focus of future work, we hope that this additional insight may provide more precise guidance as to where classical probability calculus is appropriate and not in the context of quantum mechanics.

Note that HS do not claim that all interpretations must adhere to their framework, stating (\cite{Harrigan:2010}, p. 134), ``Therefore, to categorize any given interpretation of the quantum formalism as $\psi$-ontic or $\psi$-epistemic, it is first necessary to cast it into the mold of an ontological model. If an interpretation resists being so cast, then it cannot be fit into our categorization.'' The result is that a successful interpretation of quantum mechanics must resist being so cast.

\section{On interpretations}

Our analysis entails an interesting consequence for the philosophical discussion concerning the interpretation of quantum theory. Indeed, many scholars use the PBR result to vindicate the empirical adequacy of a given framework---especially those in which the quantum state is somehow supposed to represent real physical objects---or to rule out a specific reading of the quantum formalism. Here we warn physicists and philosophers that no argument resting on the Harrigan \& Spekkens approach should be employed in such metaphysical debates, given that no theory satisfying the requirements imposed by their categorization is able to reproduce \emph{all} the results of quantum mechanics.

Therefore, we not only agree with Hance, Rarity and Ladyman in saying that the HS definitions of $\psi$-ontic and $\psi$-epistemic models do not capture the actual meaning usually assigned to the terms ``ontic'' and ``epistemic'' as they are employed when interpreting quantum theories, but also say the following:
\begin{quote}
	\textbf{If a certain interpretation of the quantum formalism\footnote{As for instance de Broglie-Bohm theory, Everett's relative state formulation, relational quantum mechanics or the many-worlds perspective \emph{etc.}.} is fully empirically equivalent to quantum mechanics, then it cannot be categorized as a $\psi$-ontic model.}
\end{quote}
\noindent In turn, the logical conjunction of our result and the PBR theorem entails that no interpretation of quantum theory can be correctly categorized according to the HS approach.

Another philosophical issue affecting the definition of $\psi$-ontic model must be mentioned. In their paper Harrigan \& Spekkens require that if a model is $\psi$-ontic, the ontic state $\lambda$ fully identifies the quantum state, i.e.\ only one quantum state is compatible with the ontic state of the system. This fact, in turn, means that $\psi$ should be understood realistically, as an entity representing some real physical property of the system, or even the system itself as in $\psi$-complete models. However, as interestingly noted by Halvorson, realism towards the quantum state is highly problematic, since each representation relation proposed in the literature entails several conundrums that should be avoided---cf.\  \cite{Halvorson:2019} for details. With regard to this, the HS framework does not clearly spell out what is meant by the fact that $\psi$ ``represents'' a real property of physical systems. In other words, to claim that in $\psi$-complete models $\lambda$ is uniquely represented by a single quantum state leaves many ontological questions unanswered. For instance, it is not specified in which way $\psi$ represents the ontic state: is it a direct or indirect representation? If $\psi$ is a real object, as one would expect in $\psi$-complete models, is there a literal identification between $\psi$ and $\lambda$? On these and other issues the Harrigan \& Spekkens framework does not give guidance.

Therefore, not only is their categorization shown to be empty, but also it does not offer a rigorous philosophical ground on which to cast the interpretational debate.

\section{Conclusion}

By using standard information theoretic techniques, we have shown that $\psi$-ontic models are not compatible with quantum information theory, and therefore with quantum mechanics itself. Given that the PBR theorem already ruled out $\psi$-epistemic models, the HS categorization of ontological models is found to be fundamentally empty. Thus, such a framework should be employed neither to classify quantum interpretations, nor to draw conclusions on the nature of the quantum state.

The key problem of the HS categorization is that it is ``not quantum enough''. The framework in fact tries to hide the non-classicality at the level of each ontic state which, by construction, cannot be decomposed further using a standard probability measure. However, HS definitions entail that statistical mixtures are simply classical mixtures in the sense that they are captured by a standard probability measure. But quantum mixtures are non-classical in the same way that the quantum transition probabilities defined by the inner product are non-classical.\footnote{The convex space formed by the mixtures is radically different \cite{Bengtsson:2017}. There is also debate as to whether proper and improper mixtures are distinguishable in the context of quantum mechanics \cite{Kirkpatrick:2001}, \cite{DEspagnat:2001}.} Therefore, the HS categorization prevents one from describing quantum statistical mixing.

Our conclusion is that the PBR theorem finds that $\psi$-epistemic models are untenable not because there is a problem with epistemic models in particular, but rather there is an underlying problem in the HS definitions. Specifically, the problem of the HS categorization is the use of standard probability theory in composite systems. If understood in this manner, the PBR result is the first part of a more general argument showing the inadequacy of the HS framework to categorize quantum theories, an argument that we have completed with the proposed no-go theorem against $\psi$-ontic models. The direct consequence of this is that the HS categorization should not be employed as a tool to classify quantum interpretations or to draw philosophically sound conclusions on the nature of quantum mechanics.

As a final note, one may argue that our result cannot possibly be correct because there are actual examples of $\psi$-ontic models. To our knowledge, however, none of them is able to reproduce all results from quantum statistical mechanics (e.g. derive the Boltzmann distribution from entropy maximization) or quantum information theory (e.g. derive the Holevo bound). Since these are derived from the more general framework of quantum mechanics, one cannot cherry pick which results a model has to reproduce to be a valid model of quantum theory: \emph{all} results must be correctly reproduced.

\section{Acknowledgements}

Andrea Oldofredi is grateful to the Funda\c{c}$\tilde{\mathrm{a}}$o para a Ci\^encia e a Tecnologia (FCT) for financial support (Grant no. 2020.02858.CEECIND).  This work is in connection to Assumptions of Physics, a larger project that aims to identify a handful of physical principles from which the basic laws can be rigorously derived  (\url{https://assumptionsofphysics.org}).

\bibliography{bibliography}
\clearpage

\section*{Appendix: calculation}
\label{A}
\textbf{Entropy of mixed non-overlapping distributions}. We want to show that, given two non-overlapping probability distributions $\rho_1$ and $\rho_2$, the entropy of $\rho = \frac{1}{2} \rho_1 + \frac{1}{2} \rho_2$ is given by
\begin{equation}
	H(\rho) = 1 + \frac{1}{2} H(\rho_1) + \frac{1}{2} H(\rho_2). \tag{\ref{entropy_nonoverlap}}
\end{equation}

Let $U_1, U_2 \subset \Lambda$ be the respective supports of the distributions, which are non-overlapping i.e.\ $U_1 \cap U_2 = \emptyset$. We then have
\begin{align*}
	H(\rho) &= - \int_\Lambda \rho \log \rho d\lambda \\
	&= -\int_{U_1} \rho \log \rho d\lambda -\int_{U_2} \rho \log \rho d\lambda \\
	&= -\int_{U_1} \frac{1}{2} \rho_1 \log \frac{1}{2} \rho_1 d\lambda -\int_{U_2} \frac{1}{2} \rho_2 \log \frac{1}{2} \rho_2 d\lambda \\
	&= - \frac{1}{2} \int_{U_1} \rho_1 \log \frac{1}{2} d\lambda - \frac{1}{2} \int_{U_1} \rho_1 \log \rho_1 d\lambda \\
	&- \frac{1}{2} \int_{U_2} \rho_2 \log \frac{1}{2} d\lambda - \frac{1}{2} \int_{U_2} \rho_2 \log \rho_2 d\lambda \\
	&= - \frac{1}{2} \log \frac{1}{2} - \frac{1}{2} \log \frac{1}{2} + \frac{1}{2} H(\rho_1) + \frac{1}{2} H(\rho_2) \\
	&= 1 + \frac{1}{2} H(\rho_1) + \frac{1}{2} H(\rho_2). \\
\end{align*}

\textbf{Entropy of quantum mixed states}. We want to show that, given two states $\psi$ and $\phi$, the entropy of the mixed state $\rho = \frac{1}{2}|\psi\rangle\langle\psi| + \frac{1}{2}|\phi\rangle\langle\phi|$ is
\begin{equation}\label{entropy}
	H(\rho) = H\left(\frac{1+|\langle\psi|\phi\rangle|}{2}, \frac{1-|\langle\psi|\phi\rangle|}{2}\right).
\end{equation}

\begin{center}
	\begin{tikzpicture}[scale = 1]
		\draw (0,0) circle (2);
		\node at (-2.3,0) {$-$};
		\node at (2.3,0) {$+$};
		\node at (2,1.4) {$\psi$};
		\node at (2,-1.4) {$\phi$};
		\draw (-2,0) -- (2,0);
		\begin{scope}
			\clip(0,0) circle (2);
			\draw (0,0) -- (2,1.4);
			\draw (0,0) -- (2,-1.4);
			\draw (1.634,1.4) -- (1.634,-1.4);
		\end{scope}
		\fill (1.634,0) circle (0.05);
		\node at (1.8,.2) {$\rho$};
	\end{tikzpicture}
\end{center}

Note that $\psi$ and $\phi$ will identify a two-dimensional subspace which can be thought, without loss of generality, as a qubit and therefore can be represented by a Bloch sphere. The picture represents the intersection of the Bloch sphere with the plane identified by $\psi$ and $\phi$. As $\rho$ is an equal mixture of the two states, it will be represented by the midpoint between the two. Taking the line that goes through $\rho$ and the center of the sphere, we can see that $\rho$ can also be seen as the mixture of the states $+$ and $-$ which, since they represent equal and opposite directions, form a basis. To diagonalize $\rho$, then, means to express it in terms of $+$ and $-$.

If $\theta_{\psi\phi}$ is the angle between $\psi$ and $\phi$, we have
\begin{equation}
	|\langle \psi | \phi \rangle |^2 = \cos^2 \frac{\theta_{\psi\phi}}{2}.
\end{equation}
The angle is divided by two because the angle on the Bloch sphere (i.e. in physical space) is double the angle in the Hilbert space. For example, for $z^+$ and $z^-$ the angle on the Bloch sphere would be $\pi$ and the inner product is zero (i.e. opposite directions in physical space correspond to orthogonal states).

Now we express $\psi$ and $\phi$ in terms of $+$ and $-$, remembering that they form a basis. Given that $\rho$ is at the midpoint, the figure is vertically symmetric. The angle between $\psi$ and $+$, then, is half of $\theta_{\psi\phi}$. The inner product between $\psi$ and $+$ is
\begin{equation}
	\begin{aligned}
		|\langle \psi | + \rangle |^2 &= \cos^2 \frac{\theta_{\psi +}}{2} \\
		&= \cos^2 \frac{\theta_{\psi\phi}}{4}.
	\end{aligned}
\end{equation}
Keeping in mind that we are composing vectors in the Hilbert space (and not in the geometry of the physical space) we have
\begin{align*}
	\left|\psi\right>&=\cos\frac{\theta_{\psi\phi}}{4}\left|+\right>+\sin\frac{\theta_{\psi\phi}}{4}\left|-\right> \\
	\left|\phi\right>&=\cos\frac{\theta_{\psi\phi}}{4}\left|+\right>-\sin\frac{\theta_{\psi\phi}}{4}\left|-\right>.
\end{align*}

The density matrices corresponding to the pure states are
\begin{align*}
	\left|\psi\right>\left<\psi\right|&=\cos^2\frac{\theta_{\psi\phi}}{4}\left|+\right>\left<+\right|\\
	&+\cos\frac{\theta_{\psi\phi}}{4}\sin\frac{\theta_{\psi\phi}}{4}\left(\left|+\right>\left<-\right|+\left|-\right>\left<+\right|\right) \\
	&+\sin^2\frac{\theta_{\psi\phi}}{4}\left|-\right>\left<-\right| \\
	\left|\phi\right>\left<\phi\right|&=\cos^2\frac{\theta_{\psi\phi}}{4}\left|+\right>\left<+\right|\\
	&-\cos\frac{\theta_{\psi\phi}}{4}\sin\frac{\theta_{\psi\phi}}{4}\left(\left|+\right>\left<-\right|+\left|-\right>\left<+\right|\right) \\
	&+\sin^2\frac{\theta_{\psi\phi}}{4}\left|-\right>\left<-\right|.
\end{align*}

We can now calculate the mixture
\begin{align*}
	\frac{1}{2}(|\psi\rangle\langle\psi| &+ |\phi\rangle\langle\phi|) \\
	&=\cos^2\frac{\theta_{\psi\phi}}{4}\left|+\right>\left<+\right| +\sin^2\frac{\theta_{\psi\phi}}{4}\left|-\right>\left<-\right| \\
	&=\frac{1+\cos\frac{\theta_{\psi\phi}}{2}}{2}\left|+\right>\left<+\right| +\frac{1-\cos\frac{\theta_{\psi\phi}}{2}}{2}\left|-\right>\left<-\right| \\
	&=\frac{1+|\langle\psi|\phi\rangle|}{2}\left|+\right>\left<+\right| +\frac{1-|\langle\psi|\phi\rangle|}{2}\left|-\right>\left<-\right|. \\
\end{align*}

As $\rho$ is in a diagonal form, the entropy is given by \ref{entropy}.

\end{document}